\documentstyle[aps,prb,epsfig,twocolumn]{revtex}
\draft
\begin{document}
\title{Boron Isotope Effect in Superconducting MgB$_2$.}
\author{S. L. Bud'ko, G. Lapertot\thanks{On leave from 
Commissariat a l'Energie Atomique, 
DRFMC-SPSMS, 38054 Grenoble, France}, C. Petrovic, 
C. E. Cunningham\thanks{On leave from Dept. of Physics, 
Grinnell College, Grinnell, IA 50112}, 
N. Anderson, and   P. C. Canfield}
\address{Ames Laboratory and  Department of Physics and Astronomy 
Iowa State University, Ames, IA  50011}
\date{\today}
\maketitle
\begin{abstract}
We report the preparation method of, and boron isotope effect for 
MgB$_2$, a new 
binary intermetallic superconductor with a remarkably high 
superconducting transition 
temperature $T_c$($^{10}$B) = 40.2 K.  Measurements of 
both temperature dependent magnetization and specific heat 
reveal a 1.0 K shift in $T_c$ between Mg$^{11}$B$_2$ and 
Mg$^{10}$B$_2$. Whereas such a high 
transition temperature might 
imply exotic coupling mechanisms, the boron isotope effect in MgB$_2$
is consistent with 
the material being a phonon-mediated BCS superconductor.\\
\end{abstract} 


The discovery of superconductivity with $T_c \approx$ 39K in 
magnesium diboride (MgB$_2$) 
was announced in January 2001 \cite{Jpn}. It caused excitement 
in the solid state physics 
community because it introduced a new, simple (3 atoms per unit cell) 
binary intermetallic 
superconductor with a record high (by almost a factor of two) 
superconducting transition 
temperature for a non-oxide and non-C$_{60}$-based compound. 
The reported value of $T_c$ 
seems to be either above or at the limit suggested theoretically 
several decades ago for 
BCS, phonon-mediated superconductivity \cite {McMil,Max}. 
An immediate 
question raised by this discovery is whether this 
remarkably high $T_c$ is due to some 
form of exotic coupling.  Therefore, any experimental data that 
can shed light on the 
mechanism of superconductivity in this material are of 
keen interest. 

One probe of the extent to which phonons mediate 
superconductivity is the 
isotope effect \cite{isoRev1,isoRev2}. 
In the classical form of BCS theory \cite{BCS}, 
the isotope coefficient $\alpha$, defined by the relation
$T_c \propto M^{-\alpha}$, where M is the mass of the 
element, is equal to 1/2. For simple metals like Hg, 
Pb, Sn, and Zn, 
the isotope coefficient is 
found experimentally to be close to 1/2. More detailed and 
realistic theories predict 
slight deviations from $\alpha$ = 1/2 \cite{McMil,Carbotte}.
In this Letter, we describe how to prepare high-quality 
powders of MgB$_2$ and, more importantly, present data on the 
boron isotope effect, which 
is consistent with phonon mediated coupling within the 
framework of the BCS model.

\begin{figure}
\epsfxsize=0.9\hsize
\vbox{
\centerline{
\epsffile{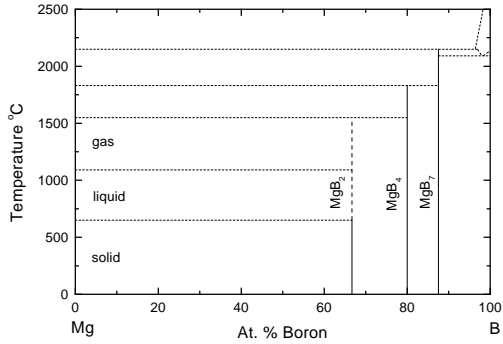}
}}
\caption{Proposed, schematic, binary phase diagram for B -– Mg system 
(After Ref. [9]).}
\label{PhDia}
\end{figure}

MgB$_2$ crystallizes in the hexagonal AlB$_2$ type 
structure, which consists 
of alternating hexagonal layers of Mg atoms and graphite-like 
honeycomb layers of B 
atoms. This material, along with other $3d-5d$ transition metal 
diborides, has been studied for 
several decades, mainly as a promising technological material 
\cite{Borides}.  The B -– Mg binary 
phase diagram \cite{PhD} is shown in Fig. 1.  As can be seen, 
MgB$_2$ decomposes 
peritectically and 
has no exposed liquid-–solidus line.  Whereas the growth of 
single crystals of this 
compound promises to be a difficult problem, high quality 
powders can be formed in the 
following manner.  Elemental Mg (99.9 \% pure in lump form) 
and isotopically pure boron 
(99.5 + \% pure, $<$ 100 mesh) are combined in a sealed Ta tube in a 
stoichiometric ratio.  
The Ta tube is then sealed in a quartz ampoule, placed in a 
950$^{\circ}$ C box furnace for two 
hours, and then removed and allowed to cool to room temperature.  
The quartz ampoule 
and the Ta tubing are not attacked, but there is a distinct 
bowing out of the Ta tube where 
the MgB$_2$ powders form.  The height of this bowing scales 
with the height of the MgB$_2$ 
powders and seems to be associated with an expansion of the B 
powder as the MgB$_2$ 
forms, rather than a Mg vapor pressure (which 
would bow out the 
tube over its whole length).  It should be noted that if larger 
pieces of B are used, then 
this reaction scheme does not produce homogeneous material.  
This, combined with the 
phase diagram shown in Fig. \ref{PhDia}, implies that, at least 
to some extent, the reaction takes 
place via the diffusion of Mg into the B particles.

\begin{figure}
\epsfxsize=0.9\hsize
\vbox{
\centerline{
\epsffile{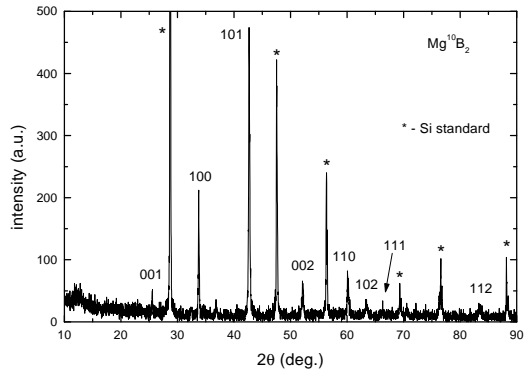}
}}
\caption{Powder X-ray (Cu K$\alpha$ radiation)
diffraction spectra of Mg$^{10}$B$_2$ 
(with h,k,l values) and Si standard (*). }
\label{XRD}
\end{figure}

The powder X-ray diffraction pattern of the Mg$^{10}$B$_2$ powder 
is shown in Fig. \ref{XRD} 
with the peaks indexed to the hexagonal unit cell of MgB$_2$.  
From Fig. \ref{XRD} the unit cell 
lattice parameters for Mg$^{10}$B$_2$ are 
a = 3.1432 $\pm$ 0.0315 
and c = 3.5193 $\pm$ 0.0323 \AA. 
 
The temperature dependent magnetization of the samples was measured in a 
Quantum Design 
MPMS-7 SQUID magnetometer in an applied field of 25G.  
An onset criterion of 2\% of 
the full, low temperature diamagnetic signal was used to determine 
T$_c$ from the zero-field-cooled $M(T)$ data that were 
taken on warming.  In this letter we report measurements on 
four types of samples with the following 
morphologies:  Mg$^{10}$B$_2$, Mg$^{11}$B$_2$ and 
Mg$^{10}$B$^{11}$B were 
all solid pieces of sample cut 
from the pellet that formed in the Ta reaction tubes whereas, 
the commercial sample was a 
fine powder.  The three solid, isotopic samples each 
had $M/H > 150\%$ of $-1/4\pi$ at low 
temperature and the powder sample had $M/H > 200$\% of $-1/4\pi$.  
By assuming spherical or slightly plate-like grains both of these 
values are consistent with demagnetization effects and 
100\% diamagnetism. The plotted 
magnetization data are all normalized to $–1$ at low temperatures 
for easier comparison. The specific heat data were taken using the 
heat capacity option of a Quantum Design PPMS-9 system in zero 
and 90 kG applied field.

\begin{figure}
\epsfxsize=0.9\hsize
\vbox{
\centerline{
\epsffile{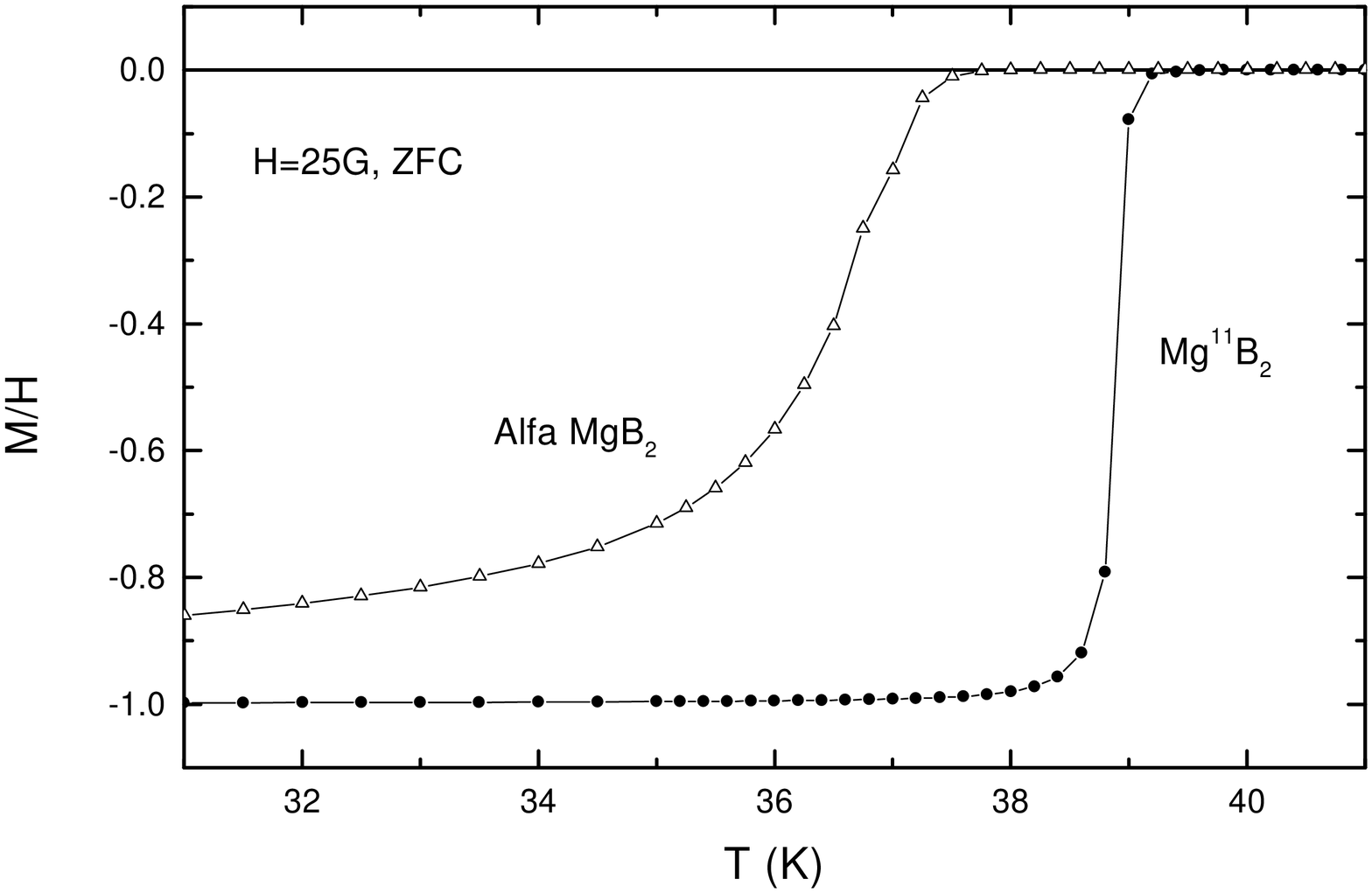}
}}
\caption{Magnetization divided by applied field as a 
function of temperature for 
Mg$^{11}$B$_2$ and natural boron sample of MgB$_2$ from 
Alfa-Aesar.  Data are normalized to $–1$
at 5 K, as discussed in text.}
\label{B11Alfa}
\end{figure}

The magnetization curves of a commercial (Alfa-Aesar 98\% pure) 
MgB$_2$ powder and prepared Mg$^{11}$B$_2$ material are shown in 
Fig. \ref{B11Alfa}. The commercial powder has a
lower T$_c$ (37.5 K) and a broader superconducting transition.  
At present, it is not clear what 
causes the suppression of $T_c$, but the difference may be due 
to impurities present in the material \cite{Alfa}. 

\begin{figure}
\epsfxsize=0.9\hsize
\vbox{
\centerline{
\epsffile{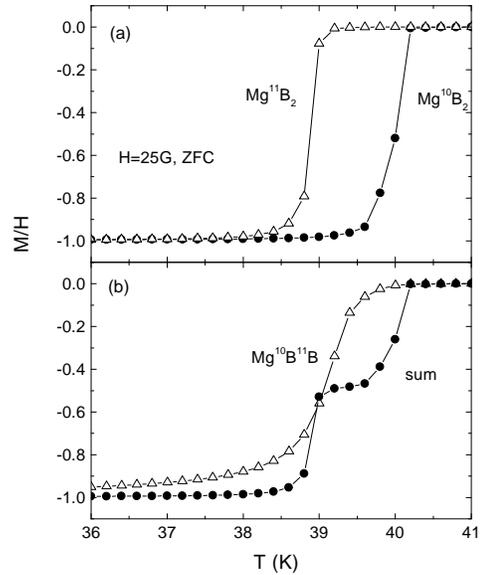}
}}
\caption{(a) Magnetization divided by applied field as a 
function of temperature for 
Mg$^{10}$B$_2$ and Mg$^{11}$B$_2$.  (b) Magnetization divided 
by applied field as a function of 
temperature for Mg$^{10}$B$^{11}$B and sum of Mg$^{10}$B$_2$
and Mg$^{11}$B$_2$ data shown in panel (a).  
Data are normalized to $-1$ at 5 K, as discussed in text.}
\label{F4}
\end{figure}

Figure \ref{F4}a presents the temperature-dependent magnetization 
of Mg$^{10}$B$_2$ and Mg$^{11}$B$_2$.  
There is a clear separation between the data of Mg$^{10}$B$_2$ and 
Mg$^{11}$B$_2$.  Using the 2\% onset 
of diamagnetism criterion mentioned above, the superconducting 
transition temperatures 
are 39.2K for Mg$^{11}$B$_2$ and 40.2K for Mg$^{10}$B$_2$.  
The widths
of the transitions (90\% -- 10\%) 
are 0.4 K and 0.5 K for Mg$^{11}$B$_2$ and Mg$^{10}$B$_2$, 
respectively.

\begin{figure}
\epsfxsize=0.9\hsize
\vbox{
\centerline{
\epsffile{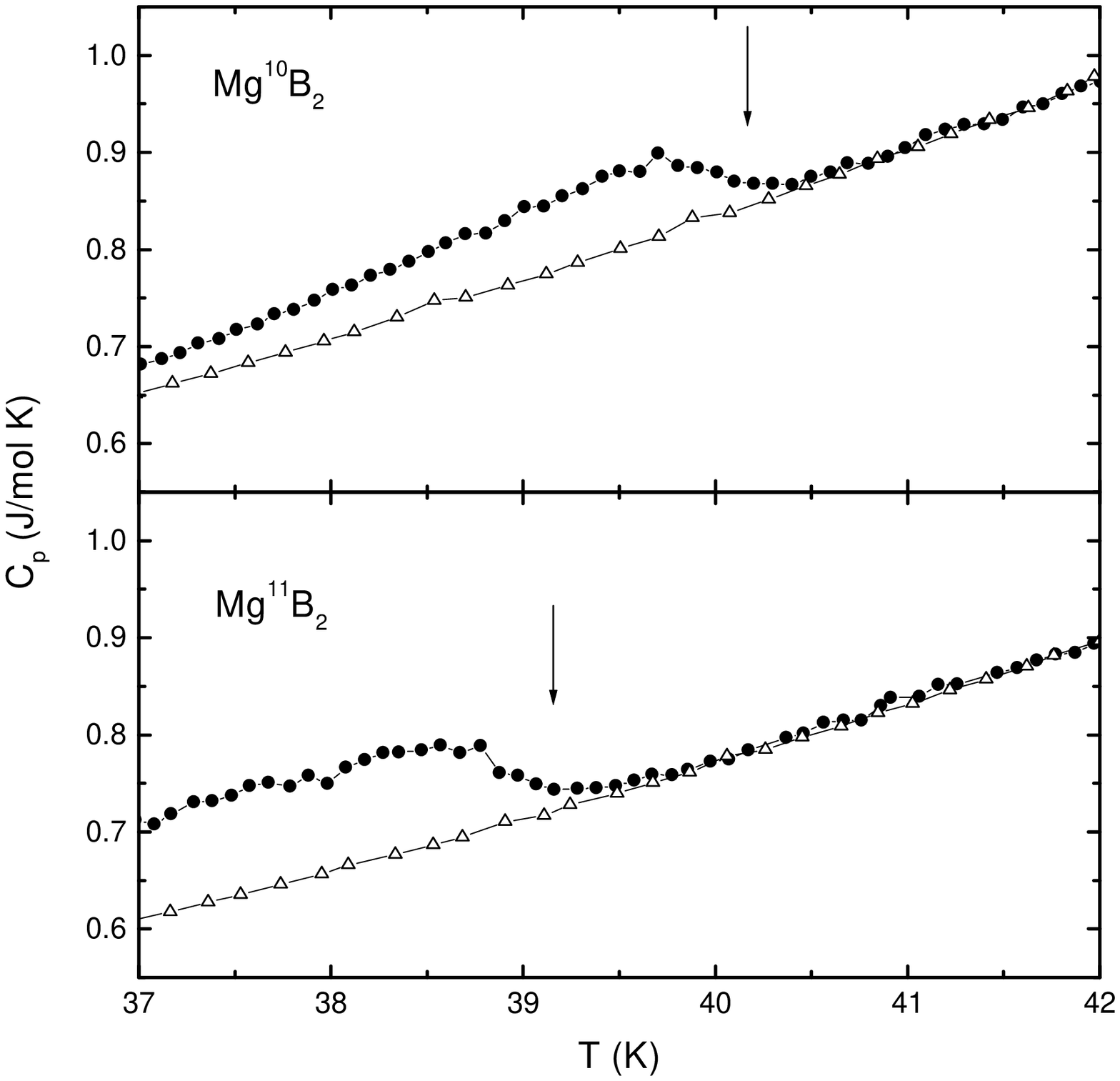}
}}
\caption{Temperature dependent specific heat of Mg$^{10}$B$_2$ 
and Mg$^{11}$B$_2$ in
zero (filled circles) and 90 kG (open triangles) applied field
for temperatures near the $T_c$. 
Arrows mark transition temperatures  
determined from the magnetization measurements shown in Fig. 4a.}
\label{HC}
\end{figure}

In Fig. \ref{HC} the temperature-dependent specific heat data 
for Mg$^{10}$B$_2$ and Mg$^{11}$B$_2$ in zero and 90 kG applied 
field are shown. Whereas the data were collected between 2 K and 50 K, 
Fig. \ref{HC} presents a more limited temperature range to 
clearly show the shift in $T_c$ associated with the isotope effect. 
The transitions are shifted by the same 1.0 K seen in Fig. \ref{F4}a. 
Due to the temperature of the superconducting transition and the 
porous nature of the samples it is difficult to extract precise 
values of the Debye temperature and the linear coefficient of 
the specific heat from these data, but we can estimate 
$\Theta_D$ = 750 $\pm$ 30 K and $\gamma$ = 3 $\pm$ 1 mJ/mol K$^2$. 
This leads to values of $\Delta$$C_p$/$\gamma$$T_c$ near 
unity. The primary point of these data should not be lost though: 
a clear isotope shift in $T_c$ can be seen in both magnetization 
and specific heat data.

To understand this substantial shift in $T_c$, it should be 
noted that if $T_c$ is assumed 
to scale with the square root of the formula unit mass, 
then the shift in $T_c$ would be expected 
to be 0.87 K.  On the other hand, if $T_c$ is assumed to scale 
with the square root of just the 
boron mass (i.e. the $T_c$ is mediated solely by boron vibrations), 
then the shift in $T_c$ would be 
1.9 K.  These two values of $\Delta$$T_c$ provide a caliper 
of the size of the isotope effect within the simpliest 
BCS framework. Viewed in this light, the shift in 
$T_c$ of 1.0 K implies that 
the phonon modes mediating the 
superconductivity are somewhat weighted toward being 
boron-like in character.

More formally, we can estimate the partial (boron) isotope 
exponent $\alpha_B$ 
in this compound via $\alpha_B = - {\Delta} \ln T_c/{\Delta} \ln M_B$
\cite{isoRev1,isoRev2}. From the measured 
values of $T_c$, the boron isotope exponent can be estimated
as $\alpha_B = 0.26 \pm 0.03$. It is worth 
mentioning that this value is close to the boron isotope 
exponents obtained for the YNi$_2$B$_2$C 
and LuNi$_2$B$_2$C borocarbides, \cite{isopoly,isosingle} 
where theoretical work \cite{Mattheiss} suggested that the 
phonons responsible for the superconductivity are 
high-frequency boron A$_{1g}$ optical 
modes.  Early band-structure calculations \cite{oldband} 
suggested substantial electron transfer from 
the magnesium atom to the two boron atoms in the unit cell.   
Recent band structure work 
\cite{Vova} suggests that the superconductivity in MgB$_2$ is 
essentially due to the 
metallic nature of the boron sheets.

Figure \ref{F4}b presents data on a 50-50 mixture of boron isotopes: 
Mg$^{10}$B$^{11}$B.  Also 
shown is the normalized sum of the pure Mg$^{10}$B$_2$ 
and Mg$^{11}$B$_2$ magnetization data shown in 
Fig. \ref{F4}a.  Given that the 
starting materials were a lump of Mg and grains of 
$^{10}$B and $^{11}$B, 
if there were no mixing 
between the boron particles as the MgB$_2$ was formed then one 
might expect the data to 
look like the sum plot, i.e. separate grains of 
Mg$^{10}$B$_2$ and Mg$^{11}$B$_2$ acting independently.  
As can be seen, the Mg$^{10}$B$^{11}$B data does 
not show two steps and 
manifests a significantly broadened transition.  
Using the 2\% criterion, $T_c$ = 39.9 K but, 
more significantly, the width of the transition is
2.1 K, a factor of four 
broader than either of the isotopically 
pure samples.  It should be noted that the Mg$^{10}$B$^{11}$B sample 
was made from exactly the 
same starting chemicals, via the same technique, and had the 
same morphology (a 
sintered lump) as the two isotopically pure samples.  
The origin of this broadening is not as of yet 
clear, but it may hint that 
the effect of isotopic disorder on the boron phonon modes 
plays a significant role. 

In conclusion, a significant boron isotope effect 
($\Delta T_c$ = 1.0 K, 
partial isotope 
exponent $\alpha_B \approx$ 0.26) was observed in MgB$_2$. 
This shift is clearly seen in both magnetization 
and specific heat measurements. This observation is 
consistent with a phonon 
mediated BCS superconducting mechanism in this compound 
and with the possibility that 
boron phonon modes are playing an important role.
\\

We appreciate useful discussions with Doug Finnemore, R. W. McCallum, K. 
Dennis and V. Antropov. 
Ames Laboratory is operated for the US Department of Energy by Iowa 
State University under Contract No. W-7405-Eng-82. This work
was supported by the Director for Energy Research, Office of 
Basic Energy Sciences.


\begin{references}
\bibitem{Jpn}J. Akimitsu, Symposium on Transition Metal
Oxides. Sendai, January 10, 2001; J.Nagamatsu, N.Nakagawa, 
T.Muranaka, Y.Zenitani, and J.Akimitsu, unpublished.
\bibitem{McMil}W. L. McMillan, Phys. Rev. {\bf 167}, 331 (1968).
\bibitem{Max}see for example E. G. Maksimov in: 
Superconductivity, Proc. P. N. Lebedev Institute v. 86, ed. 
N. G. Basov (Consultants Bureau, NY, London, 1977) p.105.
\bibitem{isoRev1}J. P. Franck in: Physical Properties of High 
Temperature Superconductors IV, ed. D. M. Ginsberg  
(World Scientific, Singapore, 1994) p.189.
\bibitem{isoRev2}R. Kishore in: Studies of High 
Temperature Superconductors Volume 29, ed. A. Narlikar  
(Nova Science Publishers, Commack, NY, 1999) p.23.
\bibitem{BCS}J. Bardeen, L. N. Cooper, and J. R. Schrieffer, 
Phys. Rev. {\bf 108}, 1175 (1957).
\bibitem{Carbotte}for review see: J. P. Carbotte, Rev. Mod. Phys.
{\bf 62}, 1027 (1990).
\bibitem{Borides}see for example: B. Aronsson, T. Lundstrom, and
S. Rundquist, Borides, Silisides and Phosphides (Methuem, London, 
1965); J. L. Hoard, and R. E. Hughes, The Chemistry of Boron and Its 
Compounds (J. Wiley, NY, 1967).
\bibitem{PhD}Binary Alloy Phase Diagrams, Second Edition, 
Edited by T. Massalski, (A.S.M International, 1990).
\bibitem{Alfa}Alfa Aesar, A Johnson Matthey Company, stock \# 88149:
Magnesium boride, 98\% (assay) MgB$_2$ (possible impurities are not 
specified). It should be noted that the powder X-ray diffraction 
(Cu K$\alpha$ radiation) 
on this sample shows an extra unidentified peak at $2\theta \approx 
15.4^\circ$.
\bibitem{isopoly}D. D. Lawrie, and J. P. Franck, Physica C, 
{\bf 245}, 159 (1995). 
\bibitem{isosingle}K. O. Cheon, I. R. Fisher, and P. C. Canfield, 
Physica C, {\bf 312}, 35 (1999).
\bibitem{Mattheiss}L. F. Mattheiss, T. Siegrist, and R. J. Cava, 
Solid State Commun. {\bf 91}, 587 (1994).
\bibitem{oldband}D. R. Armstrong, and P. G. Perkins, 
J. Chem. Soc., Faraday Trans. 2, {\bf 75}, 12 (1979).
\bibitem{Vova}J. Kortus, I. I. Mazin. K. D. Belashchenko, 
V. P. Antropov, and L. L. Boyer, cond-mat/0101446.
\end{references}
\end{document}